\documentclass[%
 reprint,
showpacs,
 amsmath,amssymb,
 aps,
 pra,
]{revtex4-1}

\usepackage[cmyk]{xcolor}

\usepackage{graphicx} 
\usepackage{dcolumn} 
\usepackage{bm} 

\usepackage[utf8]{inputenc}
\usepackage[T1]{fontenc}
\usepackage{times}
\usepackage{flafter} 
\usepackage{textgreek} 
\usepackage{mathrsfs}

\newcommand{\pipi}{\text{\textpi}}

\newcommand{\ii}{\mathrm{i}}
\newcommand{\ee}{\mathrm{e}}

\renewcommand{\Re}{\operatorname{Re}}
\renewcommand{\Im}{\operatorname{Im}}

\DeclareMathOperator{\Tr}{Tr}
\DeclareMathOperator{\diag}{diag}

\newcommand{\kronecker}{\operatorname{\delta}}

\newcommand{\qstate}{\mathit{\Psi}}
\newcommand{\spacestate}{\mathit{\Psi}}
\newcommand{\spinstate}{\mathit{\Psi}}

\newcommand{\velel}{{v_\mathrm{ee}}}
\newcommand{\gaMM}{\mathit{\Gamma}}

\newcommand{\singlet}{{(\mathrm S)}}
\newcommand{\triplet}{{(\mathrm T)}}

\newcommand{\ntilde}[1]{\langle\tilde #1|\tilde #1\rangle}
\newcommand{\nntilde}[2]{\langle\tilde #1|\tilde #2\rangle}

\newcommand{\pr}{^\prime}
\newcommand{\conj}{^\ast}

\newcommand{\adj}{^\dag}
\newcommand{\trans}{^\mathrm{T}}
\newcommand*\diff{\mathop{}\!\mathrm{d}}
\newcommand{\derivative}{\partial}
\newcommand{\odd}{\text{ odd}}
\newcommand{\even}{\text{ even}}

\makeindex

\begin{document}

\title{Equations of motion for natural orbitals of strongly driven two-electron systems}

\author{J.\ Rapp}
\affiliation{%
 Institut für Physik, Universität Rostock, 18051 Rostock, Germany
}
\author{M.\ Brics}
\affiliation{%
 Institut für Physik, Universität Rostock, 18051 Rostock, Germany
}
\author{D.\ Bauer}%
\thanks{Corresponding author: dieter.bauer@uni-rostock.de}
\affiliation{%
 Institut für Physik, Universität Rostock, 18051 Rostock, Germany
}

\date{\today}

\begin{abstract}
Natural orbital theory is a computationally useful approach to the few and many-body quantum problem. While natural orbitals are known and applied since many years in electronic structure applications, their potential for time-dependent problems is being investigated only since recently. Correlated two-particle systems are  of particular importance because the structure of the two-body reduced density matrix expanded in natural orbitals is known exactly in this case. However, in the time-dependent case the natural orbitals carry time-dependent phases that allow for certain time-dependent gauge transformations of the first kind. Different phase conventions will, in general, lead to different equations of motion for the natural orbitals. A particular phase choice allows us to derive the exact equations of motion for the  natural orbitals of any (laser-) driven two-electron system {\em explicitly}, i.e., without any dependence on quantities that, in practice, require further approximations. For illustration, we solve the  equations of motion for a model helium system. Besides calculating the spin-singlet and spin-triplet ground states, we show that the linear response spectra and the results for resonant Rabi flopping are in excellent agreement with the benchmark results obtained from the exact solution of the time-dependent Schr\"odinger equation. 
\end{abstract}

\maketitle

\section{Introduction}
$N$-electron systems in full dimensionality that are strongly driven by, e.g., an intense laser field, can be simulated on an {\em ab initio} time-dependent Schr\"odinger equation (TDSE)-level only up to $N=2$ (see, e.g., \cite{scrinzi}). This embarrassingly small 
number calls for efficient time-dependent ``even-not-so-many''-body quantum approaches that are applicable beyond linear response.

In order to overcome the unpleasant exponential complexity scaling 
of a correlated many-particle state $|\qstate(t)\rangle$, quantities of less dimensionality should be used \cite{coulson}.
An example for such an approach is  time-dependent density functional theory (TDDFT). The 
Runge-Gross theorem of TDDFT \cite{runge-gross,ullrich} ensures that  the single-particle density 
$n(\vec r, t)$ is, in principle,  sufficient to calculate all observables of a time-dependent 
many-body quantum system. However, the---principally exact---equations of motion (EOM) of TDDFT 
for the auxiliary Kohn-Sham orbitals  involve a generally unknown exchange-correlation (XC) 
functional. It has been shown that the {\em non-adiabaticity} of the XC functional is essential for 
the description of correlated dynamics~\cite{helbig}. However, essentially all practicable approximations 
to the unknown exact XC functional neglect memory effects but make use of the numerically 
strongly favorable adiabatic approximation. But even if the exact single-particle density 
$n(\vec r, t)$ was reproduced there remains  the problem of extracting the relevant observables 
from $n(\vec r, t)$ in practice. For instance, it is unknown how multiple ionization probabilities, 
photoelectron spectra, let alone differential and correlated ones, can be {\em explicitly} calculated from  
$n(\vec r, t)$ alone \cite{petersilka,wilken,wilken2}.

Because of these practical difficulties with  $n(\vec r, t)$-based TDDFT it is an obvious idea to use 
less reduced quantities as building bricks, e.g.,  reduced density matrices (or quantities related to them; see, for instance, \cite{colemanyukalov,cios,gido,mazz,mazz2,pernalTD,appelthesis,gies1,giesbthesis,requist,appelgross}).  In fact, the knowledge of the two-body reduced density matrix (2-RDM)  is sufficient to explicitly calculate any observable involving one and two-body operators.
However, as density matrices are still high-dimensional objects it is not attractive to solve the EOM for them directly. 
L\"owdin introduced so-called natural orbitals (NOs) and occupation numbers (ONs) as eigenfunctions and eigenvalues of the one-body reduced density matrix (1-RDM), respectively \cite{loewdin1}, and investigated the stationary two-electron case in great detail  \cite{loewdintwoelecs}. NOs have the same dimensionality as single-particle wavefunctions and may be used as basis functions for configuration interaction (CI) approaches, for instance. In fact, one may hope that NOs form the best possible basis set with respect to some measure, e.g., $\|\Psi-\Phi\|^2$, where $\Phi$ is a CI approximation to the exact wavefunction  $\Psi$. 
 Recently, it has been shown  that this is true only  for special cases  (including two electrons), and how NOs may be used to generate the best basis~\cite{giesbnobasis2014}.

In the current paper we derive the general EOM for NOs renormalized to the corresponding ONs [called {\em time-dependent renormalized natural orbital theory} (TDRNOT)] before we specialize on the time-dependent two-body problem. For the interacting two-body system the structure of the 2-RDM expressed in terms of NOs is exactly known but unique only up to certain combinations of time-dependent NO phases.  Different NO phase choices will lead to different EOM. For a particular phase choice  \cite{giesbthesis} the 2-RDM depends only on the time-dependent ONs and NOs but not on additional time-dependent phases, and the TDRNOT Hamiltonian in the EOM is thus  {\em exactly} and {\em explicitly} known.
 Hence,  solving the EOM for the NOs is equivalent to  the solution of the corresponding TDSE. In particular, the $N$-representability (also called ``quantum marginal'')  problem (see, e.g., \cite{colemanyukalov}) is not an issue in this simplest time-dependent few-body case.

In practice we wish (and  need) to truncate the number of NOs we take into account, which introduces propagation errors in the numerical solution of the TDRNOT EOM. We therefore benchmark our approach with  a system for which we can actually solve the TDSE numerically exactly: the widely used (laser-) driven one-dimensional helium model atom (see, e.g., \cite{model1,model2}).  It has already been shown in \cite{tdrnot} that our approach---even with a ground-state ``frozen'' effective Hamiltonian---covers highly-correlated phenomena such as double excitations and autoionization, both inaccessible by practicable, adiabatic  TDDFT~\cite{krueger}. The frozen-Hamiltonian calculations (also known as the ``bare'' response) was used in \cite{tdrnot} because with the phase convention chosen there the time-evolution of the above-mentioned phases, and thus the consistent time-evolution of the 2-RDM, was unknown.

\medskip

The paper is organized as follows. The basic theory of reduced density matrices and NOs regarding two-electron systems is introduced in section~\ref{sec:theory}. The new phase convention is introduced in section~\ref{subsec:newphaseconv}, the respective EOM for the NOs is discussed in section~\ref{sec:discussing-new-phase-convention}. Finally, we benchmark the performance of TDRNOT in section~\ref{sec:results}, before we conclude and give an outlook in section~\ref{sec:conclout}. Some of the derivations and details are given in   appendices~\ref{appendix:start}-\ref{appendix:end}.

\section{Two-body natural orbital theory }\label{sec:theory}
Atomic units (a.u.) are used throughout. In some cases, operator hats are used to emphasize the non-diagonality of an operator in some particular space.

\subsection{Density matrices, natural orbitals, and occupation numbers}
Starting point in the case of a two-body system is the pure two-body density matrix (2-DM) 
\begin{align}
  \hat\gamma_2(t)
    &=
      |\qstate(t)\rangle
      \langle\qstate(t)|. \label{eq:2dm}
\end{align}
The 1-RDM $\hat\gamma_1(t)$ then reads
\begin{align}
  \hat\gamma_1(t)
    &=
      \sum_{i=1}^2\Tr_i
      \hat\gamma_2(t)
    =
      2\Tr_1
      \hat\gamma_2(t)
    =
      2\Tr_2
      \hat\gamma_2(t) \label{eq:def-1rdm}
\end{align}
where the partial trace $\Tr_i$ means tracing out all degrees of freedom of particle $i$.  Both $\hat\gamma_2(t)$ and $\hat\gamma_1(t)$ are Hermitian.

The NOs $|k(t)\rangle$ and ONs $n_k(t)$ are defined as eigenstates and eigenvalues of the 1-RDM, respectively,
\begin{align}
  \hat\gamma_1(t)
  |k(t)\rangle
    &=
      n_k(t)
      |k(t)\rangle.
\end{align}
As $\hat\gamma_1(t)$ is Hermitian, the $n_k(t)$ are real, and the $|k(t)\rangle$ are orthogonal. We further assume the $|k(t)\rangle$ to be normalized to unity so that $\{|k(t)\rangle\}$ is a complete, orthonormal basis. With this convention, the spectral decomposition of the 1-RDM reads
\begin{align}
  \hat\gamma_1(t)
    &=
      \sum_{k=1}^\infty
      n_k(t)
      |k(t)\rangle
      \langle k(t)|. \label{eq:1rdm-spectral-decomposition}
\end{align}
Because of the normalization of the two-particle state
$
  \langle\qstate(t)|\qstate(t)\rangle
    =
      1$
we have
$
  \Tr\hat\gamma_2(t)
    =
      1$ and
  $
  \Tr\hat\gamma_1(t)
    = N=
      2$,
where $N=2$ arises as the number of particles in the system, and $\Tr$ without subscript is understood as the trace over whatever degrees of freedom the operator to be traced has. Evaluating the trace of $\hat\gamma_1(t)$ leads to
\begin{align}
  \sum_k
  n_k(t)
    &= N=
      2. \label{eq:on-sum}
\end{align}

The 2-DM can be expanded in NOs as well,
\begin{align}
  \hat\gamma_2(t)
    &=
      \sum_{ijkl}
      \gamma_{2,ijkl}(t)
      |i(t), j(t)\rangle
      \langle k(t), l(t)|, \label{eq:formal-2dm-expansion}
\end{align}
where the shorthand notation for tensor products
$
  |i(t), j(t)\rangle
    =
      |i(t)\rangle
      |j(t)\rangle
    =
      |i(t)\rangle
      \otimes
      |j(t)\rangle
$
is used, and the expansion coefficients $\gamma_{2,ijkl}(t)$ formally read
\begin{align}
  \gamma_{2,ijkl}(t)
    &=
      \langle i(t), j(t)|
      \hat\gamma_2(t)
      |k(t), l(t)\rangle. \label{eq:formal-2dm-expansion-coefficients}
\end{align}

\subsection{Renormalized natural orbitals}
In TDRNOT, renormalized natural orbitals (RNOs)
\begin{align}
  |\tilde k(t)\rangle
    &=
      \sqrt{n_k(t)}
      |k(t)\rangle,
  &
  \langle\tilde k(t)|\tilde k(t)\rangle
    &=
      n_k(t)
\end{align}
are introduced because it is numerically beneficial to store and {\em unitarily propagate} the combined quantity $|\tilde k(t)\rangle$ instead of using the coupled set of equations for $|k(t)\rangle$ and $n_k(t)$ \cite{tdrnot}. In RNOs, the expansions \eqref{eq:1rdm-spectral-decomposition} and  \eqref{eq:formal-2dm-expansion}
 read
\begin{gather}
  \hat\gamma_1(t)
    =
      \sum_k
      |\tilde k(t)\rangle
      \langle\tilde k(t)|, \\
  \hat\gamma_2(t)
    =
      \sum_{ijkl}
      \tilde\gamma_{2,ijkl}(t)
      |\tilde i(t), \tilde j(t)\rangle
      \langle\tilde k(t), \tilde l(t)|,
\end{gather}
with renormalized expansion coefficients
\begin{align}
  \tilde\gamma_{2,ijkl}(t)
    &=
      \frac{
        \gamma_{2,ijkl}(t)
      }{
        \sqrt{
          n_i(t)
          n_j(t)
          n_k(t)
          n_l(t)
        }
      }. \label{eq:renormalized-2dm-expansion-coefficients}
\end{align}

\subsection{Peculiarities of the two-electron state}
Based on the exchange antisymmetry
\begin{align}
  \hat{\mathcal P}^{(1,2)}
  |\qstate(t)\rangle
    &=
      -|\qstate(t)\rangle, \label{eq:exchange}
\end{align}
any two-electron state $|\qstate(t)\rangle$ can be expanded in its RNOs $|\tilde k(t)\rangle$ as
\begin{align}
  |\qstate(t)\rangle
    &=
      \sum_{k\odd}
      \frac{
        \ee^{\ii\varphi_k(t)}
      }{
        \sqrt{2 n_k(t)}
      }
      \Big[
        |\tilde k(t),\tilde k\pr(t)\rangle
        - |\tilde k\pr(t),\tilde k(t)\rangle
      \Big] \label{eq:state-expansion}
\end{align}
with the ''prime operator`` acting on a positive integer $k$ as
\begin{align}
  k\pr
    &=
      \begin{cases}
        k+1 & \text{if $k\odd$}\\
        k-1 & \text{if $k\even$},
      \end{cases}
  &
  k
    &>
      0. \label{eq:prime-operator}
\end{align}
A proof of \eqref{eq:state-expansion} is provided in appendix~\ref{appendix:state-expansion}. The conditions
\begin{align}
  n_k(t)
    &=
      n_{k\pr}(t),
  &
  n_k(t)
    \in
      [0, 1] \label{eq:pairwise-on-degeneracy-and-interval}
\end{align}
for the ONs follow.

If we require $|\qstate(t)\rangle$ to be an eigenstate of the spin operators $\hat S^2$ and $\hat S_z$ at all times we can write
\begin{align}
  |\qstate(t)\rangle
    &=
      |\spacestate(t)\rangle_x
      \otimes
      |\spinstate\rangle_\sigma \label{eq:state-factorization}
\end{align}
where $|\spinstate\rangle_\sigma$ is a time-independent spin component and  $ |\spacestate(t)\rangle_x$ is the spatial part.
The spin part needs not to be considered explicitly as long as the Hamiltonian does not act on it. However, it affects the  exchange symmetry of $|\spacestate(t)\rangle_x$.

\subsubsection{Spin singlet}
In the spin-singlet case, 
\begin{align}
  |\spinstate\rangle_\sigma
    &=
      \frac{1}{\sqrt 2}
      \Big[
        \left|
          \uparrow
          \downarrow
        \right\rangle_\sigma
        - \left|
          \downarrow
          \uparrow
        \right\rangle_\sigma
      \Big] \label{eq:spin-singlet}
\end{align}
so that
\begin{align*}
  \hat{\mathcal P}^{(1,2)}
  |\spinstate\rangle_\sigma
    &=
      -|\spinstate\rangle_\sigma,
  &
  \hat{\mathcal P}^{(1,2)}
  |\spacestate(t)\rangle_x
    &=
      +|\spacestate(t)\rangle_x.
\end{align*}
The RNOs $|\tilde k(t)\rangle$ may be factorized
\begin{align}
  |\tilde k(t)\rangle
    &=
      |\tilde k(t)\rangle_x
      \otimes
      \begin{cases}
        \left|\uparrow\right\rangle_\sigma & \text{if $k\odd$}\\
        \left|\downarrow\right\rangle_\sigma & \text{if $k\even$}
      \end{cases} \label{eq:rno-factorization-singlet}
\end{align}
with pairwise equal components
\begin{align}
  |\tilde k(t)\rangle_x
    &=
      |\tilde k\pr(t)\rangle_x. \label{eq:singlet-spatial-equality}
\end{align}
Insertion into~\eqref{eq:state-expansion} and comparison with~\eqref{eq:state-factorization} and~\eqref{eq:spin-singlet} yields 
\begin{align}
  |\spacestate(t)\rangle_x
    &=
      \sum_{k\odd}
      \frac{
        \ee^{\ii\varphi_k(t)}
      }{
        \sqrt{n_k(t)}
      }
      |\tilde k(t), \tilde k(t)\rangle_x, \label{eq:spatial-singlet}
\end{align}
which indeed has the desired exchange symmetry.

\subsubsection{Spin triplet}\label{sec:triplet-factorization}
In the three spin-triplet cases we have
\begin{align*}
  \hat{\mathcal P}^{(1,2)}
  |\spinstate\rangle_\sigma
    &=
      +|\spinstate\rangle_\sigma,
  &
  \hat{\mathcal P}^{(1,2)}
  |\spacestate(t)\rangle_x
    &=
      -|\spacestate(t)\rangle_x.
\end{align*}
Each of the three spin-triplet configurations is associated with a different factorization of the RNOs.
Consider, e.g., 
\begin{align}
  |\spinstate\rangle_\sigma
    &=
      \left|
        \uparrow
        \uparrow
      \right\rangle_\sigma.\label{eq:spin-triplet-upup}
\end{align}
In this case we choose
\begin{align}
  |\tilde k(t)\rangle
    &=
      |\tilde k(t)\rangle_x
      \otimes
      \left|\uparrow\right\rangle_\sigma, \label{eq:rno-factorization-triplet-upup}
\end{align}
leading to the correct
\begin{align}
  |\spacestate(t)\rangle_x
    &=
      \sum_{k\odd}
      \frac{
        \ee^{\ii\varphi_k(t)}
      }{
        \sqrt{2 n_k(t)}
      }
      \Big[
        |\tilde k(t), \tilde k\pr(t)\rangle_x
        - |\tilde k\pr(t), \tilde k(t)\rangle_x
      \Big] \label{eq:spatial-triplet}
\end{align}
[without an additional condition like~\eqref{eq:singlet-spatial-equality}]. 
The structure~\eqref{eq:spatial-triplet} of $|\spacestate(t)\rangle_x$ also holds for the two remaining triplet configurations, as shown in appendix~\ref{appendix:triplet-rno-factorization}. Moreover, the RNO factorizations in spin and spatial components can be chosen such that $|\tilde k(t)\rangle_x$ is invariant when switching between the different spin triplets.

\subsection{Exact 2-DM}
The universal expansion~\eqref{eq:state-expansion} of any two-electron state $|\qstate(t)\rangle$ in terms of RNOs $|\tilde k(t)\rangle$ implies fundamental knowledge about the connection between the 2-DM $\hat\gamma_2(t)$ and the RNOs, as revealed by inserting~\eqref{eq:state-expansion} into~\eqref{eq:2dm}. 
As a result, $\tilde\gamma_{2,ijkl}(t)$ can be calculated using~\eqref{eq:formal-2dm-expansion-coefficients} and~\eqref{eq:renormalized-2dm-expansion-coefficients},
\begin{align}
  \tilde\gamma_{2,ijkl}(t)
    =
      (-1)^{i-k}
      \frac{
        \ee^{
          \ii\left[
            \varphi_i(t)
            - \varphi_k(t)
          \right]
        }
      }{
        2\sqrt{n_i(t) n_k(t)}
      }
      \kronecker_{i,j\pr}
      \kronecker_{k,l\pr}. \label{eq:2dm-renormalized-expansion-coefficients}
\end{align}
One sees that the renormalized expansion coefficients $\tilde\gamma_{2,ijkl}(t)$ are only nonvanishing for {\em paired} index combinations. Both the first index pair $\{i, j\}$ and the second index pair $\{k, l\}$ must contain one odd and one even index. Moreover, the ``distance'' between the paired indices is unity, i.e.,
\begin{align}
  |i-j|
    &=
      1,
  &
  |k-l|
    &=
      1
  &
  \text{ if }
  \tilde\gamma_{2,ijkl}(t)
    \neq
      0.
\end{align}

\subsection{Phase conventions}\label{subsec:newphaseconv}
So far, no assumption has been made concerning the phases of the NOs. Any phase transformation according to
\begin{align}
  |\underline{k}(t)\rangle
    &=
      \ee^{\ii\vartheta_k(t)}
      |k(t)\rangle \label{eq:no-phase-transformation}
\end{align}
yields a new set of NOs $\{|\underline k(t)\rangle\}$ for the same 1-RDM $\hat\gamma_1(t)$ with the same ONs $\{n_k(t)\}$. This phase freedom originates from the definition of NOs as eigenstates of $\hat\gamma_1(t)$, allowing for arbitrary time-dependent NO phases because they vanish in~\eqref{eq:1rdm-spectral-decomposition}. However, the expansion~\eqref{eq:state-expansion} of $|\qstate(t)\rangle$ requires phase factors $\ee^{\ii\varphi_k(t)}$ in order to compensate for the phase freedom in the NOs. The transformation~\eqref{eq:no-phase-transformation} thus also involves a phase transformation
\begin{align}
  \underline{\varphi_k}(t)
    &=
      \varphi_k(t)
      - \vartheta_k(t)
      - \vartheta_{k\pr}(t),
  &
  k&\odd.
\end{align}
This is in analogy of ``gauge transformations of the first kind'' in field theory. However, the TDRNOT Hamiltonian is, in general,  {\em not} invariant under such phase transformations. Observables {\em are} invariant.

In order to derive EOM for the NOs, one needs to choose well-defined NO phases. Two choices are presented in the following.

\subsubsection{Time-dependent phases} \label{sec:former-phase-convention}
In the first publication on TDRNOT~\cite{tdrnot}, the NO phases were fixed by 
\begin{align}
  \langle\underline k(t)|\derivative_t|\underline k(t)\rangle
    &=
      0, \label{eq:former-phase-convention}
\end{align}
which can formally be fulfilled by the transformation
\begin{align}
  \vartheta_k(t)
    &=
      \ii\int^t
      \langle k(t\pr)|\derivative_{t\pr}|k(t\pr)\rangle
      \diff t\pr.
\end{align}
As a result, the phases $\underline{\varphi_k}(t)$ are time-dependent, which requires the solution of coupled EOM for the NOs and $\{\underline{\varphi_k}(t)\}$ because the time evolution of the RNOs depends on these phases via $\tilde\gamma_{2,ijkl}(t)$ [see \eqref{eq:2dm-renormalized-expansion-coefficients} and the EOM in section~\ref{sec:discussing-new-phase-convention} below].

\subsubsection{Phase-including natural orbitals} \label{sec:PINO-phase-convention}
The phase freedom can be utilized to transform-away the time-dependence of $\underline{\varphi_k}(t)$. One easily verifies that, e.g., the transformation
\begin{align}
  \vartheta_k(t)
    &=
      \vartheta_{k\pr}(t)
    =
      \frac{1}{2}\left[
        \varphi_k(t)
        - \varphi_{k,0}
      \right],
  &
  k&\odd \label{eq:transformation-to-new-phases}
\end{align}
yields arbitrarily tunable constant phases $\underline{\varphi_k}(t)\equiv\varphi_{k,0}\in\mathbb{R}$. Depending on the spin configuration [singlet $\singlet$ or triplet $\triplet$] we choose the atomic He ground state phase factors
\begin{align}
  \ee^{\ii\varphi_{k,0}^\singlet}
    &=
      2\kronecker_{k,1} - 1,
  &
  \ee^{\ii\varphi_{k,0}^\triplet}
    &=
      1,
  &
  k&\odd  \label{eq:frozen-phase-factors}
\end{align}
so that a real ground state wavefunction yields real NOs in position space representation.

Based on this phase convention one may derive EOM for $|\underline k(t)\rangle$ such that all time-dependence is incorporated in the {\em phase-including} NOs (PINOs) \cite{giesbthesis,giesbgritsbaer,Meer} and the ONs. Note that the transformation~\eqref{eq:transformation-to-new-phases} does not remove all phase freedom because one can still distribute the phase between any pair $|\tilde k(t),\tilde k\pr(t)\rangle$ in the triplet case. The missing constraint is given by~\eqref{eq:fix-new-phase-convention} in the derivation of the respective EOM.

In the following we will omit the underline in $|\underline k(t)\rangle$ for the phase-including (R)NOs.

\section{Equations of motion for renormalized phase-including natural orbitals}\label{sec:discussing-new-phase-convention}
We consider a two-electron Hamiltonian
\begin{align}
  \hat H^{(1, 2)}(t)
    &=
      \hat h^{(1)}(t)
      + \hat h^{(2)}(t)
      + \velel^{(1, 2)}, \label{eq:hamiltonian}
\end{align}
where the single-particle part $\hat h(t)$ incorporates kinetic energy, binding potential, and, e.g.,  the coupling to (time-dependent) external fields, and $\velel$ is the  electron-electron interaction.  Superscripts indicate the particle indices.
The time evolution of the NOs is expanded as
\begin{align*}
  \ii\derivative_t|k(t)\rangle
    &=
      \sum_m
      \alpha_{km}(t)
      |m(t)\rangle.
\end{align*}
We see that the phase convention~\eqref{eq:former-phase-convention} chosen in~\cite{tdrnot} is equivalent to setting $\alpha_{kk}(t)\equiv 0$. Instead, for the PINO phase convention of section~\ref{sec:PINO-phase-convention} we employ the diagonal elements $\alpha_{kk}(t)$ in order to modify the EOM such that the phases $\{\varphi_k\}$ stay constant. A useful expression for $\alpha_{kk}(t)$ in terms of RNOs is derived in appendix~\ref{appendix:derivation-diagonal-alpha} for the two-electron case considered here. Adding the new contributions associated with $\alpha_{kk}(t)$ to the EOM for the RNOs derived in~\cite{tdrnot} yields (time arguments of the RNOs suppressed) 
\begin{align}
  \ii\derivative_t
  |\tilde n\rangle
    &=
      \hat h(t)
      |\tilde n\rangle
      + \mathcal A_n(t)
      |\tilde n\rangle \nonumber\\
    &\quad
      + \sum_{k\neq n} \mathcal B_{nk}(t)
      |\tilde k\rangle
      + \sum_k \hat{\mathcal C}_{nk}(t)
      |\tilde k\rangle \label{eq:new-eom}
\end{align}
with
\begin{align}
  \mathcal A_n(t)
    &=
      -\frac{1}{n_n(t)}
      \Re\sum_{jkl}
      \tilde\gamma_{2,njkl}(t)
      \langle\tilde k\tilde l|
      \velel
      |\tilde n\tilde j\rangle, \label{eq:new-coefficients-a}
\end{align}
\begin{align}
  \mathcal{B}_{nk}(t)
    &=
      \frac{2}{n_k(t)-n_n(t)}
      \sum_{jpl}\left[
        \tilde\gamma_{2,kjpl}(t)
        \langle\tilde p\tilde l|
        \velel
        |\tilde n\tilde j\rangle
      \right. \nonumber\\
    &\qquad
      \left.
        - \tilde\gamma_{2,plnj}(t)
        \langle\tilde k\tilde j|
        \velel
        |\tilde p\tilde l\rangle
      \right],\quad
  k
    \neq
      n\pr , \label{eq:coefficients-b}
\end{align}
and
\begin{align}
  \hat{\mathcal C}_{nk}(t)
    &=
      2\sum_{jl}
      \tilde\gamma_{2,kjnl}(t)
      \langle\tilde l|\velel|\tilde j\rangle. \label{eq:new-coefficients-c}
\end{align}
Only $ \mathcal A_n(t)$ is modified due to $\alpha_{kk}(t)\not\equiv 0$ whereas $\mathcal{B}_{nk}(t)$ and $\hat{\mathcal C}_{nk}(t)$ are invariant under the phase transformation.

Special treatment is required regarding the $\mathcal{B}_{nk}(t)$ of the pairs $k=n\pr$ because of the pairwise degeneracy $n_k(t)=n_{k\pr}(t)$. Recalling (A8) of~\cite{tdrnot}, 
\begin{align*}
  \alpha_{np}(t)\Big[
    n_p(t)
    -n_n(t)
  \Big]
    &=
      \Big[
        n_p(t)
        -n_n(t)
      \Big]
      \langle p|
      \hat h(t)
      |n\rangle\nonumber\\
    &\quad
      + 2\sum_{jkl}
      \gamma_{2,pjkl}(t)
      \langle kl|
      \velel
      |nj\rangle\nonumber\\
    &\quad
      - 2\sum_{jkl}
      \gamma_{2,klnj}(t)
      \langle pj|
      \velel
      |kl\rangle,
\end{align*}
it follows that $\alpha_{np}(t)$ is undetermined for $n_p(t)=n_n(t)$ so that $\mathcal{B}_{nn\pr}(t)$ cannot be obtained by following the derivation in~\cite{tdrnot}. This reflects the fact that, independent of the choice of phase, eigenstates corresponding to degenerate eigenvalues are not uniquely defined. In terms of NOs one finds that $ |\underline{k}\rangle$, $|\underline{k\pr}\rangle$ according
\begin{align}
  \left(\begin{array}{c}
    |\underline{k}\rangle\\
    |\underline{k\pr}\rangle
  \end{array}\right)
    &=
      \left(\begin{array}{cc}
        \cos\left[\theta_k(t)\right] & \sin\left[\theta_k(t)\right]\\
        -\sin\left[\theta_k(t)\right] & \cos\left[\theta_k(t)\right]
      \end{array}\right)
      \left(\begin{array}{c}
        |k\rangle\\
        |k\pr\rangle
      \end{array}\right) \label{eq:superposition-for-degeneracies}
\end{align}
yield the same state $|\qstate(t)\rangle$ for any choice of $\{\theta_k(t)\}$. In practice, this is not an issue for the spin singlet because the additional freedom is removed by the particular choice of the product ansatz~\eqref{eq:rno-factorization-singlet}. 
For the spin triplet we choose $\alpha_{nn\pr}(t)\equiv 0$. Hence, we replace the corresponding coefficients $\mathcal{B}_{nn\pr}(t)$ in the spin-triplet case by
\begin{eqnarray}
\lefteqn{  \mathcal{B}_{nn\pr}(t)  =
      - \frac{1}{n_n(t)}} \label{eq:eom-addition}\\
    &&\qquad\qquad\quad\times \left[\vphantom{\sum_{jpl}}
        \langle\tilde n\pr|\hat h(t)|\tilde n\rangle
      + 2\sum_{jpl}
        \tilde\gamma_{2,plnj}(t)
        \langle\tilde n\pr\tilde j|\velel|\tilde p\tilde l\rangle
      \right]. \nonumber
\end{eqnarray}

\subsection{Occupation numbers during imaginary-time propagation}\label{sec:discussion-full-imaginary-time-propagation}
It has already been shown~\cite{tdrnot} that the spin-singlet ground state is a stationary point of the EOM when propagating the RNOs in imaginary time. Unfortunately, using the phase convention of section~\ref{sec:former-phase-convention} used in \cite{tdrnot}, the ONs are invariant during imaginary-time propagation. As a consequence, one needs to inject the correct ONs for the ground state. A useful criterion for the ground state configuration $\{n_k\}$ can be derived by means of variational calculus minimizing the total energy $E\geq E_0$. In this work we supplement the variational calculus with an additional constraint for finding the spin-triplet ground state. Details are given in appendix~\ref{appendix:triplet-variation}. The result for the orbital energies reads
\begin{align*}
  \epsilon_k
    &=
      \frac{1}{n_k}
      \left[
        \langle\tilde k|\hat h_0|\tilde k\rangle
        + \sum_{ijl}\tilde\gamma_{2,ijkl}
        \langle\tilde k\tilde l|\velel|\tilde i\tilde j\rangle
      \right].
\end{align*}
The ONs in the ground state configuration have to be such that 
\begin{align}
  E
    &=
      \epsilon_k
      + \epsilon_{k\pr}, \label{orbital-energies-criterion}
\end{align}
i.e., each sum of two associated orbital energies in the ground state equals the total energy $E$. For the spin-singlet ground state all orbital energies are equal, i.e.,
$
  \epsilon_k^{(\mathrm{S})}
    =
      \epsilon^{(\mathrm{S})}$.
In the spin-triplet case, one additional Lagrange parameter $\epsilon^\mathrm{d}_k$ for odd $k$ is introduced to ensure that $n_k=n_{k\pr}$. Because of $\epsilon^\mathrm{d}_k$, individual triplet orbital energies are generally not equal,
\begin{align*}
  \epsilon_k^{(\mathrm{T})}
    &=
      \epsilon^{(\mathrm{T})}
      + \epsilon^\mathrm{d}_k\kronecker_{k\odd}
      - \epsilon^\mathrm{d}_{k-1}\kronecker_{k\even}.
\end{align*}

Using the phase convention of section~\ref{sec:former-phase-convention} one may tune the ONs $n_k$ such that the orbital energies $\epsilon_k$ fulfill~\eqref{orbital-energies-criterion} when the RNOs $|\tilde k\rangle$ are converged to the stationary point of the imaginary-time propagation. For more than two NOs per electron this is a multidimensional problem so that a Newton-Raphson scheme may be employed to find the correct ground state ONs. Details are given in appendix~\ref{appendix:newton-derivation}.

Fortunately, using the PINO phase convention of section~\ref{sec:PINO-phase-convention} simplifies the ground state search because the ONs are {\em not} constant during imaginary time propagation but adjust themselves. In fact, $\derivative_t n_n(t)$ can be calculated using
\begin{align}
  \derivative_t
  n_n(t)
    &=
      \Big[
        \derivative_t
        \langle\tilde n|
      \Big]
      |\tilde n\rangle
      + \langle\tilde n|
      \Big[
        \derivative_t
        |\tilde n\rangle
      \Big]. \label{eq:on-change}
\end{align}
Replacing $\ii\derivative_t|\tilde n\rangle$ by $-\derivative_t|\tilde n\rangle$ on the left-hand side of the EOM~\eqref{eq:new-eom} one may insert the result and its adjoint into~\eqref{eq:on-change} to obtain
\begin{align*}
  \derivative_t
  n_n(t)
    &=
      -2 n_n(t) \epsilon_n(t)
\end{align*}
for real NOs. We conclude that in the desired ground state configuration, the relative change of ONs is constant for each associated orbital pair, i.e.,
\begin{align*}
  \frac{
    \derivative_t
    \left[
      n_n(t)
      + n_{n\pr}(t)
    \right]
  }
  {
    n_n(t)
    + n_{n\pr}(t)
  }
    &=
      -E_0.
\end{align*}
As a result, the set of ground state ONs is a stationary point of the imaginary-time propagation if the restrictions~\eqref{eq:on-sum} and~\eqref{eq:pairwise-on-degeneracy-and-interval} are enforced after each timestep. In practice we find that the ONs converge to this stationary point when propagating in imaginary time. No additional criterion such as~\eqref{orbital-energies-criterion} needs to be applied for finding the ground state via imaginary-time propagation with the PINO phase convention.

\subsection{Conservation of occupation-number degeneracies}\label{sec:on-degeneracy-conservation}
Let us check whether the pairwise degeneracy of ONs~\eqref{eq:pairwise-on-degeneracy-and-interval} is conserved when propagating the RNOs in real time. As the  pairwise degeneracy results from the exchange antisymmetry, a violation of the ON degeneracies would imply a violation of the Fermionic character of the electrons described. In the actual numerical implementation we use an absorbing potential, i.e., $\hat h\adj(t)\neq\hat h(t)$, in order to remove orbital probability density approaching the grid boundaries. One then finds (suppressing time arguments of the RNOs again)
\begin{align*}
  \derivative_t n_k(t)
    &=
      2\Im\langle\tilde k|\hat h(t)|\tilde k\rangle
      + 4\Im\sum_{ijl}
      \tilde\gamma_{2,ijkl}(t)
      \langle\tilde k\tilde l|\velel|\tilde i\tilde j\rangle.
\end{align*}
If the time propagation is performed fully self-consistently, i.e., without freezing the effective Hamiltonian, and absorption is negligible, 
\begin{align*}
  \derivative_t\left[
    n_k(t)
    - n_{k\pr}(t)
  \right]
    &=
      0,
\end{align*}
as can be shown by making use of the special structure~\eqref{eq:2dm-renormalized-expansion-coefficients} of $\tilde\gamma_{2,ijkl}(t)$ in the case of two electrons.

If the absorbing potential significantly influences the ONs, the condition for the conservation of degeneracies reads
\begin{align}
  \Im\langle\underline{\tilde k}|
  \hat h(t)
  |\underline{\tilde k}\rangle
    &=
      \Im\langle\underline{\tilde k\pr}|
      \hat h(t)
      |\underline{\tilde k\pr}\rangle. \label{eq:absorption-criterion-triplet-superposition}
\end{align}
In the singlet case, \eqref{eq:absorption-criterion-triplet-superposition} always holds because the spatial components of the RNOs $|\tilde k(t)\rangle$ and $|\tilde k\pr(t)\rangle$ are equal due to the factorization~\eqref{eq:rno-factorization-singlet}. In the triplet case, there is the freedom to use superpositions \eqref{eq:superposition-for-degeneracies} such that \eqref{eq:absorption-criterion-triplet-superposition} is fulfilled for all $k$. However, in this paper we do not show results where a significant 
amount of probability density was absorbed so that the application of 
criterion~\eqref{eq:absorption-criterion-triplet-superposition} was not necessary.

\section{Results}\label{sec:results}
Results are obtained for the one-dimensional helium model atom \cite{model1,model2} described by  the Hamiltonian \eqref{eq:hamiltonian} with
\begin{align*}
  \hat h(t)
    &=
      \frac{\hat p^2}{2}
      - \frac{2}{\sqrt{x^2 + 1}}
      + A(t)\hat p,
\end{align*}
\begin{align*}
  \velel^{(1, 2)}
    &=
      \frac{1}{\sqrt{\left(x^{(1)}-x^{(2)}\right)^2 + 1}}.
\end{align*}
The interaction with an external (laser) field in dipole approximation is incorporated in velocity gauge via the vector potential $A(t)$, with the purely time-dependent $A^2$-term transformed away. 
Numerical results are shown for both the spin singlet and the spin triplet. As a first check, we confirm in section~\ref{sec:results-ground-state} that the EOM for the renormalized PINOs~\eqref{eq:new-eom}-\eqref{eq:new-coefficients-c},~\eqref{eq:eom-addition} yield the exact ground state energy and correct ONs if enough RNOs are included in the propagation. The second step is to employ the PINO EOM for a propagation in real time in order to evaluate the advantages of the PINO phase convention over the previously used  \cite{tdrnot} phase convention of section \ref{sec:former-phase-convention}. For this purpose, linear response spectra considering a different number of RNOs are discussed in section~\ref{sec:results-linres}. Rabi oscillations, as a prime example for  highly resonant and nonperturbative phenomena  that bring quantum systems far away from their ground state, are investigated in section~\ref{sec:results-rabi}.

\bigskip

In practice, the number of RNOs is truncated in order to allow for a numerical treatment. In the following, $N_\mathrm{o}$ denotes the number of spin orbitals so that $N_\mathrm{o}$ RNOs correspond to $N_\mathrm{o}/2$ different spatial orbitals for the spin singlet and $N_\mathrm{o}$  different spatial orbitals for the spin triplet. Computational details are given in \cite{tdrnot}.

\subsection{Ground state calculations}\label{sec:results-ground-state}
The ground state is obtained via imaginary-time propagation, as discussed in section~\ref{sec:discussion-full-imaginary-time-propagation}. 
Both phase conventions yield the same ground state configurations so that we do not need to distinguish between the two in this section.

The total energy and the dominant ONs for both the spin-singlet and the spin-triplet ground state are presented in Table~\ref{tab:ground-state}. TDRNOT results for different $N_\mathrm{o}$ are compared to the exact TDSE results. All TDRNOT results clearly converge to the corresponding exact TDSE value for increasing $N_\mathrm{o}$.

\newcommand{\converged}[1]{\text{\underline{$#1$}}}
\newcommand{\unconverged}[1]{#1}

\begin{table}[htbp]
  \caption{Total energy and ONs of the spin-singlet and spin-triplet ground state, respectively. Exact results obtained from the direct solution of the TDSE are compared to TDRNOT results using different $N_\mathrm{o}$. Converged digits are underlined.  }

\medskip

  \begin{tabular}{ c | c | c | c | c } \hline\hline
    Number $N_\mathrm{o}$ & Total energy & \multicolumn{3}{c}{Dominant occupation numbers}\\[0pt]
    of RNOs & $E_0$ (a.u.) & $n_1$ & $n_3/10^{-3}$ & $n_5/10^{-5}$\\\hline
    \multicolumn{5}{c}{Spin singlet}\\\hline
    $2$ (TDHF)      & $\converged{-2.2}\unconverged{24318}$ & $\unconverged{1.0000000}$  &          &\\
    $4$ (TDRNOT)    & $\converged{-2.23}\unconverged{6595}$ & $\converged{0.99}\unconverged{12665}$  & $\converged{8}.\unconverged{7335}$ &\\
    $6$ (TDRNOT)    & $\converged{-2.238}\unconverged{203}$ & $\converged{0.9909}\unconverged{590}$  & $\converged{8.3}\unconverged{142}$ & $\converged{7}\unconverged{2.683}$\\
    $8$ (TDRNOT)    & $\converged{-2.2383}\unconverged{24}$ & $\converged{0.99094}\unconverged{38}$  & $\converged{8.3}\unconverged{221}$ & $\converged{70}.\unconverged{229}$\\
    $\infty$ (TDSE) & $\converged{-2.238368}$                   & $\converged{0.9909473}$                    & $\converged{8.3053}$ & $\converged{70.744}$\\
    \hline\multicolumn{5}{c}{Spin triplet}\\\hline
    $2$ (TDHF)          & $\converged{-1.81}\unconverged{20524}$  & $\unconverged{1.00000000}$  &                             &\\
    $4$ (TDRNOT)        & $\converged{-1.816}\unconverged{0798}$  & $\converged{0.9976}\unconverged{4048}$  & $\converged{2.3}\unconverged{5952}$ &\\
    $6$ (TDRNOT)        & $\converged{-1.8161}\unconverged{870}$  & $\converged{0.99760}\unconverged{705}$  & $\converged{2.36}\unconverged{464}$ & $\converged{2}.\unconverged{8298}$\\
    $8$ (TDRNOT)        & $\converged{-1.81619}\unconverged{45}$  & $\converged{0.997606}\unconverged{56}$  & $\converged{2.362}\unconverged{67}$ & $\converged{2.9}\unconverged{581}$\\
    $\infty$ (TDSE)     & $\converged{-1.8161954}$                    & $\converged{0.99760677}$                    & $\converged{2.36220}$                   & $\converged{2.9610}$\\ \hline
  \end{tabular}
  \label{tab:ground-state}
\end{table}

$N_\mathrm{o}=2$ is equivalent to a time-dependent Hartree-Fock (TDHF) treatment or TDDFT in exact exchange-only approximation. Very similar results as in Table~\ref{tab:ground-state} have been reported in~\cite{mctdhf} using a multiconfigurational time-dependent Hartree-Fock (MCTDHF) approach. The strength of two-electron TDRNOT compared to two-electron MCTDHF is the choice of RNOs as a basis, which always guarantees the best approximation to the exact solution $|\qstate(t)\rangle$ for a given number of orbitals~\cite{giesbnobasis2014} {\em at all times} during real-time propagation.

\subsection{Linear response spectra}\label{sec:results-linres}
Starting from the spin-singlet or spin-triplet ground state, the vector potential is switched to a finite but small value ($A=0.0005$ was chosen for the results presented in the following), and the RNOs are propagated in real-time for $t_\mathrm{max}=1000$ with an enabled imaginary potential. The Fourier transform of the dipole expectation value then yields peaks at energy differences $E-E_0$ for all dipole-allowed transitions.

\begin{figure}[htbp]%

  \includegraphics{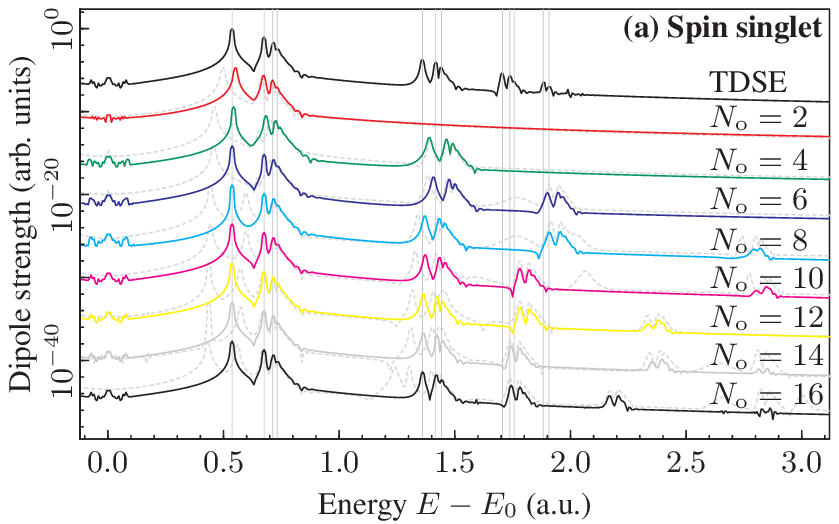}

  \includegraphics{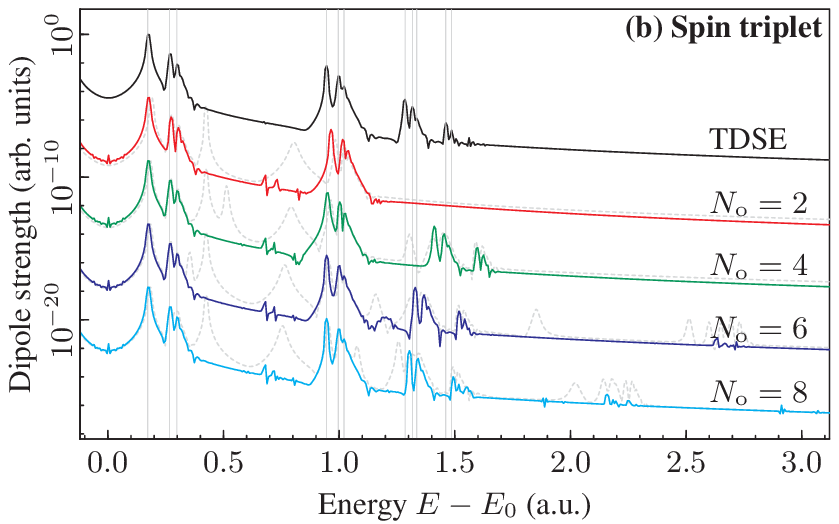}

  \caption{(Color online) Singlet (a) and triplet (b) linear response spectra for a different number of RNOs $N_\mathrm{o}$, compared to the exact TDSE result. For comparison, bare (i.e., with ground-state frozen Hamiltonian) TDRNOT results following the phase convention of section~\ref{sec:former-phase-convention} are shown with dashed lines. To guide the eyes, vertical lines indicate some of the distinct peaks in the exact TDSE spectrum.}%
 \label{fig:linres}%
\end{figure}

Figure~\ref{fig:linres} shows that the fully self-consistent TDRNOT time propagation  reproduces the exact linear response spectra (solid; labeled ``TDSE'') for both the spin singlet (a) and the spin triplet (b) if enough RNOs are taken into account. 
As already known from the bare evolution in \cite{tdrnot}, the description of doubly-excited states requires at least $N_\mathrm{o}\geq 4$ so that the ONs are not pinned to the integers $0$ or $1$.

As expected, the more series of doubly excited states are sought the more RNOs are needed. Interestingly, some peak positions of the spin singlet show an alternating convergence if one successively adds two RNOs more. For example, the peak around $E-E_0\approx1.35$ is shifted to the wrong direction from $N_\mathrm{o}=4$ to $N_\mathrm{o}=6$ but substantially shifts towards its correct position for $N_\mathrm{o}=8$. Using $N_\mathrm{o}=10$, its peak position again slightly worsens compared to the previous value whereas for $N_\mathrm{o}=12$ the energy matches almost perfectly with the TDSE peak position.

The fully self-consistent time propagation using the PINO phase convention of section~\ref{sec:PINO-phase-convention} (solid) is clearly superior to the bare evolution with the phase convention of section~\ref{sec:former-phase-convention} (dashed gray): erroneous extra-peaks are absent, and the physical peaks are shifted to the correct TDSE positions. Both effects are particularly important for more RNOs, say $N_\mathrm{o}\geq 6$. Especially for the triplet, the full propagation with PINOs leads to much better results. The bare evolution generates erroneous extra peaks for any number of RNOs, corresponding to artificial states with nondegenerate ONs. Since degenerate ONs are a consequence of the exchange antisymmetry those peaks indicate the breaking of the exchange symmetry by the bare time evolution with  the ground-state frozen Hamiltonian. This deficiency is removed by the full propagation using PINOs, as discussed in section~\ref{sec:on-degeneracy-conservation}.

MCTDHF linear response spectra for the same model have been obtained in~\cite{mctdhf}. Our Fig.~\ref{fig:linres}(a) can be directly compared with Fig.~3 there, where artificial extra peaks just above the first ionization threshold are seen. The reason for the erroneous peaks in the MCTDHF results is unknown to us. The superior performance of our TDRNOT approach using PINOs is presumably due to the  built-in optimal choice of basis set functions {\em at all times}.

It is to be  expected that our promising results translate to 3D two-electron systems. In fact, in Refs.~\cite{giesbgritsbaer,Meer} it has been shown already that only a few of the highest occupied PINOs are sufficient to capture accurately the lowest excitations in the response of the 3D two-electron systems H$_2$ and HeH$^+$.  

\subsection{Rabi oscillations}\label{sec:results-rabi}
Linear response spectra are not enough to study strong-field laser-matter interaction phenomena, which, by definition, are non-perturbative in nature and rely on electron dynamics far away from the ground state.
A prime example for non-perturbative laser-matter coupling is Rabi oscillations. 
It has been shown that Rabi oscillations are not captured within ``standard'' TDDFT~\cite{rabi-1} but that  XC functionals with memory, i.e., XC functionals beyond the adiabatic approximation, are required~\cite{helbig}. It is important to understand that adiabatic TDDFT applied to Rabi oscillations may reproduce a reasonably looking  position expectation value as a function of time~\cite{rabi-1} even though the time-dependent density is \emph{not} properly described, especially at times of population inversion, e.g., after a $\pi$-pulse. 
Instead, the ONs $n_k(t)$ as a function of time are very sensitive entities, which we use for benchmarking our TDRNOT approach via a comparison with the exact TDSE result.

We consider a Rabi oscillation between the spin-singlet ground state and the first excited state, driven by a laser of resonant frequency $\omega=0.5337$. The vector potential amplitude $A=0.0125$ of the flat-top part is linearly ramped-up over four periods. 
Propagating eight different spatial NOs, we have  $N_\mathrm{o}=16$. Due to the pairwise degeneracy follows     $n_1(t)=n_2(t), \dots, n_{15}(t)=n_{16}(t)$ so that it is sufficient to discuss $n_{k\odd}(t)$.

\begin{figure}[htbp]%

  \includegraphics{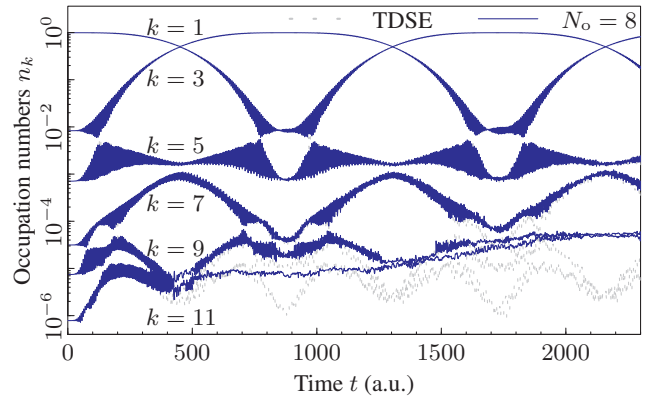}
  
  \caption{(Color online) ONs $n_k(t)$ vs time $t$ for the spin singlet in a laser field of frequency $\omega=0.5337$ resonantly tuned to the first excited state. The four most significant ONs $n_1(t), n_3(t), n_5(t), n_7(t)$ obtained by TDRNOT with $N_\mathrm{o}=16$ RNOs (solid) correctly reproduce more than two Rabi cycles of the exact TDSE propagation (dotted). Due to the truncation to a finite number of RNOs in TDRNOT, less significant orbitals are missing the proper coupling to lower orbitals, leading to erroneous behavior of small ONs over time. For longer propagation times also higher ONs are affected because the RNOs are coupled.}%
 \label{fig:rabi}%
\end{figure}

The six most significant ONs $n_1(t), n_3(t), \dots, n_{11}(t)$ predicted by the TDRNOT propagation (solid) are compared with the exact TDSE result (dotted) in Fig.~\ref{fig:rabi}. 

\subsubsection{Truncation problem}
Thanks to the proper ground state description reported in section~\ref{sec:results-ground-state}, all TDRNOT ONs start on top of the exact TDSE reference for $t=0$ in Fig.~\ref{fig:rabi}. However, already for small times $0<t\lesssim 200$ ONs $n_{13}(t)$ and $n_{15}(t)$ (not shown) begin to deviate from the correct value. Instead of the periodic oscillation with the Rabi period $2\pipi/\mathit{\Omega}_\mathrm{R}\approx 850$ and a modulation on the timescale of the laser period $2\pipi/\omega\approx 11.8$ they just approach their respective ``upper neighbor'' NO's ON. The next ONs $n_{11}(t)$ and $n_9(t)$ become quantitatively distinguishable from their respective TDSE values around $t\gtrsim 400$ and $t\gtrsim 800$. After two Rabi cycles, i.e., $t\gtrsim 1700$ also their qualitative behavior is completely wrong, showing no oscillation on the Rabi timescale any longer. Around that time $t\approx 1700$ the next higher ON $n_7(t)$ is affected and shows some small quantitative differences compared to the exact solution, although it regains the proper behavior at later times.

The origin of these imperfections regarding the least significant orbitals in the propagation lies in the truncation to a finite number $N_\mathrm{o}=16$ of RNOs taken into account. 
The EOM in section~\ref{sec:discussing-new-phase-convention} have been derived for an infinite number of coupled RNOs. It turns out that the orbital coupling via $\mathcal B_{nk}(t)$ is particularly strong for orbitals with nearby ONs so that the truncation of the orbitals $\{|\tilde{17}(t)\rangle, |\tilde{18}(t)\rangle, \dots\}$ is most severe for the least significant orbitals. Once their dynamics is spoiled, the truncation error subsequently propagates ``upwards'' due to the coupling to the respective next higher orbitals.

\subsubsection{Overall performance}
The four most significant ONs $n_1(t)$, $n_3(t)$, $n_5(t)$, $n_7(t)$  in Fig.~\ref{fig:rabi} are in a striking agreement with the exact TDSE result. Their dynamics during more than two Rabi cycles, i.e., a time period of 2300 atomic units in total, is well-described. 
Overall, the ``well-behaved'' RNOs represent more than $99.9\,\%$ of the 1-RDM so that the significant part of the Rabi dynamics is captured by TDRNOT.

The remarkable gain of TDRNOT compared to, e.g., TDDFT is that---despite the (numerically strongly favorable) locality in time---TDRNOT is capable of describing the highly resonant dynamics of Rabi oscillations. In fact, the exact  two-electron TDRNOT EOM are strictly memory-free.


\section{Conclusion and outlook} \label{sec:conclout}
In the current work, we have extended the previously introduced \cite{tdrnot} time-dependent 
renormalized natural orbital theory (TDRNOT).
We have derived the equations of motion for renormalized natural orbitals, employing 
the phase convention in which the entire time-dependence is carried by the natural orbitals themselves. 
In the two-particle case, this allows to obtain the exact equations of motion, without making any assumptions about
(or any approximations to) the 
expansion of the time-dependent two-body density matrix in natural orbitals.
As an example, we have solved the equations of motion  for a widely used  helium model atom.  In practical calculations, 
the number of natural orbitals taken into account should be as small as possible. As a truncation of the number of
natural orbitals introduces numerical errors, we have benchmarked our results by the  corresponding exact solutions of
the time-dependent Schr\"odinger  equation. Excellent agreement has been found for the spin-singlet and spin-triplet 
ground states (obtained via imaginary-time propagation), linear response spectra, and Rabi flopping dynamics
(as an example for a strongly non-perturbative, resonant phenomenon).

We are mainly interested in laser-driven {\em few}-body correlated quantum dynamics.
Besides Rabi flopping, we are currently applying the TDRNOT method successfully to other (strong-field) scenarios  
where ``standard'' time-dependent density functional theory with practicable exchange-correlation 
potentials is known to fail, e.g., nonsequential double ionization. Moreover, we are investigating the structure of the exact expansion coefficients  $\tilde\gamma_{2,ijkl}$ for three-electron systems in order to derive useful expressions that can be used to propagate the respective natural orbitals using TDRNOT.


\section*{Acknowledgment}
Fruitful discussions with M.\ Lein are acknowledged.
This work was supported by the SFB 652 of the German Science Foundation (DFG).


\appendix

\section{Expansion of a   two-fermion state in RNOs}\label{appendix:start} \label{appendix:state-expansion}
\newcommand{\basfunc}{\psi}
\newcommand{\natorb}{\phi}
Let the expansion of a two-fermion state $|\qstate(t)\rangle$ in orthonormal single-particle basis functions $|\basfunc_i(t)\rangle$ comprising spin and spatial degrees of freedom be
\begin{align*}
  |\qstate(t)\rangle
    &=
      \sum_{ij}
      \qstate_{ij}(t)
      |\basfunc_i(t), \basfunc_j(t)\rangle, \\
  \qstate_{ij}(t)
    &=
      \langle\basfunc_i(t), \basfunc_j(t)|\qstate(t)\rangle.
\end{align*}
Defining a matrix $
  \bm\qstate
    =
      \left[
        \qstate_{ij}(t)
      \right]
$ of expansion coefficients $\qstate_{ij}(t)$, the exchange antisymmetry can be expressed as $
  \bm\qstate\trans
    =
      -\bm\qstate
$. With
\begin{align*}
  \bm\basfunc
    &=
      \begin{pmatrix}
        |\basfunc_1(t)\rangle\\
        |\basfunc_2(t)\rangle\\
        \vdots
      \end{pmatrix},
  &
  \bm\basfunc\conj
    &=
      \begin{pmatrix}
        \langle\basfunc_1(t)|\\
        \langle\basfunc_2(t)|\\
        \vdots
      \end{pmatrix}, \\
  \bm\basfunc\trans
    &=
      \begin{pmatrix}
        |\basfunc_1(t)\rangle,
        |\basfunc_2(t)\rangle,
        \dots
      \end{pmatrix},
  &
  \bm\basfunc\adj
    &=
      \begin{pmatrix}
        \langle\basfunc_1(t)|,
        \langle\basfunc_2(t)|,
        \dots
      \end{pmatrix}
\end{align*}
such that
\begin{align*}
  \bm\basfunc\conj
  \bm\basfunc\adj
    &=
      \begin{pmatrix}
        \langle\basfunc_1(t)|\\
        \langle\basfunc_2(t)|\\
        \vdots
      \end{pmatrix}
      \begin{pmatrix}
        \langle\basfunc_1(t)|,
        \langle\basfunc_2(t)|,
        \dots
      \end{pmatrix} \\
    &=
      \begin{pmatrix}
        \langle\basfunc_1(t), \basfunc_1(t)| & \langle\basfunc_1(t), \basfunc_2(t)| & \dots\\
        \langle\basfunc_2(t), \basfunc_1(t)| & \langle\basfunc_2(t), \basfunc_2(t)| & \dots\\
        \vdots & \vdots & \ddots
      \end{pmatrix}
\end{align*}
the relation between a two-fermion state $|\qstate(t)\rangle$ and its coefficient matrix $\bm\qstate$ in the basis $\{|\basfunc_i(t)\rangle\}$ may be written as
\begin{align}
  \bm\qstate
    &=
      \bm\basfunc\conj
      \bm\basfunc\adj
      |\qstate(t)\rangle,
  &
  |\qstate(t)\rangle
    &=
      \bm\basfunc\trans
      \bm\qstate
      \bm\basfunc. \label{eq:relation-between-state-and-coefficient-matrix}
\end{align}
The skew-symmetric matrix $\bm\qstate=-\bm\qstate\trans$ can be factorized into unitary matrices $\bm U, \bm U\adj$ and a block-diagonal matrix $\bm\Sigma$ as~\cite[Corollary~2.6.6.~(b)]{matrixAnalysis}
\begin{align*}
  \bm\qstate
    &=
      \bm U
      \bm\Sigma
      \bm U\trans,
  &
  \bm\Sigma
    &=
      \diag(\bm\Sigma_1, \bm\Sigma_3, \bm\Sigma_5, \dots),\\
  \bm\Sigma_i
    &=
      \begin{pmatrix}
        0       & \xi_i(t) \\
        -\xi_i(t)  & 0
      \end{pmatrix},
  &
  i
    &
      \odd. 
\end{align*}
Inserting this factorization
into \eqref{eq:relation-between-state-and-coefficient-matrix}, 
one obtains an expansion 
in the transformed basis $\bm\natorb=\bm U\trans\bm\basfunc$,
\begin{align*}
  |\qstate(t)\rangle
    &=
      \bm\basfunc\trans
      \left(
        \bm U
        \bm\Sigma
        \bm U\trans
      \right)
      \bm\basfunc
    =
      \bm\natorb\trans
      \bm\Sigma
      \bm\natorb.
\end{align*}
In other words, any two-fermion state $|\qstate(t)\rangle$ can be written in the form
\begin{align}
  |\qstate(t)\rangle
    &=
      \sum_{i\odd}
      \xi_i(t)
      \Big[
        |\natorb_{i}(t), \natorb_{i\pr}(t)\rangle
        - |\natorb_{i\pr}(t), \natorb_{i}(t)\rangle
      \Big] \label{eq:qstate-in-unknown-basis}
\end{align}
where the prime operator \eqref{eq:prime-operator} was used.
Inserting~\eqref{eq:qstate-in-unknown-basis} into the 1-RDM~\eqref{eq:def-1rdm} gives
\begin{align*}
  \hat\gamma_1(t)
    &=
      \sum_{k\odd}
      2|\xi_k(t)|^2
      \Big[
        |\natorb_{k}(t)\rangle
        \langle\natorb_{k}(t)|
        + |\natorb_{k\pr}(t)\rangle
        \langle\natorb_{k\pr}(t)|
      \Big],
\end{align*}
which proves that 
$
  |k(t)\rangle
    =
      |\natorb_{k}(t)\rangle$,
i.e., the set $\{ |\natorb_{k}(t)\rangle\}$ {\em is} a set of NOs.
The corresponding eigenvalues $2|\xi_k(t)|^2$ for odd $k$, i.e., the ONs, are (at least) pairwise degenerate,
\begin{align*}
  n_k(t)
    &=
      n_{k+1}(t)
    =
      2|\xi_k(t)|^2,
  &
  k
    &
      \odd.
\end{align*}
Writing
$
  \xi_k(t)
    =
      \tfrac{1}{\sqrt 2}
      \ee^{\ii\varphi_k(t)}
      \sqrt{n_k(t)}
$ for odd $k$, and switching to RNOs $\{|\tilde k(t)\rangle\}$, a two-fermion state reads
\begin{equation}
  |\qstate(t)\rangle
    =
      \sum_{k\odd}
      \frac{
        \ee^{\ii\varphi_k(t)}
      }{
        \sqrt{2 n_k(t)}
      }
      \Big[
        |\tilde k(t),\tilde k\pr(t)\rangle
        - |\tilde k\pr(t),\tilde k(t)\rangle
      \Big].
\end{equation}


\section{Factorization of RNOs in the triplet cases}\label{appendix:triplet-rno-factorization}
Section~\ref{sec:triplet-factorization} contains a brief discussion of the very simple RNO factorization for the spin-triplet configuration $|\spinstate\rangle_\sigma=\left|\uparrow\uparrow\right\rangle_\sigma$. The case $|\spinstate\rangle_\sigma=\left|\downarrow\downarrow\right\rangle_\sigma$ is analogous.  
The factorization of the NOs for the spin triplet 
\begin{align}
  |\spinstate\rangle_\sigma
    &=
      \frac{1}{\sqrt 2}
      \left[
        \left|
          \uparrow
          \downarrow
        \right\rangle_\sigma
        + \left|
          \downarrow
          \uparrow
        \right\rangle_\sigma
      \right] \label{eq:spin-triplet-updown}
\end{align}
is more involved.
Considering both positive and negative indices $k$ one may define RNOs 
\begin{align*}
  |\tilde k(t)\rangle
    &=
      \begin{cases}
        |\tilde k(t)\rangle_x & \text{if $k>0$}\\
        |-\tilde k(t)\rangle_x & \text{if $k<0$}
      \end{cases}
      \otimes
      \begin{cases}
        \left|\uparrow\right\rangle_\sigma & \text{if $k>0$, $k\odd$}\\
        \left|\downarrow\right\rangle_\sigma & \text{if $k>0$, $k\even$}\\
        \left|\downarrow\right\rangle_\sigma & \text{if $k<0$, $k\odd$}\\
        \left|\uparrow\right\rangle_\sigma & \text{if $k<0$, $k\even$}
      \end{cases},
\end{align*} 
and a generalized prime operator acting on nonzero integer numbers $k$ according
\begin{align*}
  k\pr
    &=
      \begin{cases}
        k+1 & \text{if $k>0$, $k\odd$}\\
        k-1 & \text{if $k>0$, $k\even$}\\
        k-1 & \text{if $k<0$, $k\odd$}\\
        k+1 & \text{if $k<0$, $k\even$}
      \end{cases} .
\end{align*}
Insertion into~\eqref{eq:state-expansion} (where now both positive and negative $k$ have to be considered in the sum) yields, again, the same structure~\eqref{eq:spatial-triplet}  and the same $|\tilde k(t)\rangle_x$ as the other triplet configurations. If the Hamiltonian~\eqref{eq:hamiltonian} does not act  on spin degrees of freedom, as it is the case for the model He atom considered, the sole significance of the spin component of the state  $|\qstate(t)\rangle$ is its effect on the exchange symmetry of the spatial part, which is the same for each of the three triplet configurations.

\section{Derivation of $\alpha_{kk}$ for PINOs}\label{appendix:derivation-diagonal-alpha}
Writing~\eqref{eq:state-expansion} as
\begin{align}
  |\qstate(t)\rangle
    &=
      \sum_{i\odd}
      \xi_i(t)
      \left[
        |i, i\pr\rangle
        - |i\pr, i\rangle
      \right], \label{universal-expansion-xi}
  &
  \xi_i(t)
    &=
      \ee^{\ii\varphi_{i,0}}
      \sqrt{\frac{n_i(t)}{2}},
\end{align}
with the phase factors $\ee^{\ii\varphi_i}$ given by~\eqref{eq:frozen-phase-factors},
yields, upon insertion  into the right-hand-side of the  TDSE 
\begin{equation} \hat H(t)|\qstate(t)\rangle=\ii\partial_t |\qstate(t)\rangle \label{eq:TDSE} \end{equation}
\begin{align*}
  \hat H(t)|\qstate(t)\rangle
    &=
      \ii\sum_{i\odd}
      \left[
        \dot\xi_i(t)\left(
          |i,i\pr\rangle
          -|i\pr,i\rangle
        \right) \right. \\
& \qquad \left.
        +\xi_i(t)\left(
          |\dot{i},i\pr\rangle
          -|\dot{i\pr},i\rangle
          +|i,\dot{i\pr}\rangle
          -|i\pr,\dot{i}\rangle
        \right)
      \right].
\end{align*}
\begin{widetext}
Multiplying from the left by $\langle k, k\pr|$ for an odd $k$ gives
\begin{align}
  \langle k, k\pr|\hat H(t)|\qstate(t)\rangle
    &=
      \ii\dot\xi_k(t)
      + \xi_{k}(t)
      \left[\vphantom{\hat h(t)}
        \langle k|\derivative_t|k\rangle
        + \langle k\pr|\derivative_t|k\pr\rangle
      \right]=
      \ii\frac{\dot n_k(t)}{2 n_k(t)}
      \xi_k(t) + \ii\xi_k(t)
      \left[\vphantom{\hat h(t)}
        \alpha_{kk}(t)
        + \alpha_{k\pr k\pr}(t)
      \right]. \label{diagonal-alpha-singlet-rhs}
\end{align}
Insertion  of \eqref{universal-expansion-xi} into the left-hand-side of the TDSE \eqref{eq:TDSE} gives
\begin{align}
  \langle k, k\pr|\hat H(t)|\qstate(t)\rangle
    &=
      \xi_k(t)
      \left[
        \langle k|\hat h(t)|k\rangle
        + \langle k\pr|\hat h(t)|k\pr\rangle
      \right] + \sum_{i\odd}
      \xi_i(t)
      \left[\vphantom{\hat h(t)}
        \langle k, k\pr|\velel|i, i\pr\rangle
        - \langle k, k\pr|\velel|i\pr, i\rangle
      \right]. \label{diagonal-alpha-singlet-lhs}
\end{align}
Combination of~\eqref{diagonal-alpha-singlet-rhs} and~\eqref{diagonal-alpha-singlet-lhs} yields
\begin{align}
  \alpha_{kk}(t)
  + \alpha_{k\pr k\pr}(t)
    &=
      \langle k|\hat h(t)|k\rangle
      + \langle k\pr|\hat h(t)|k\pr\rangle
      + \sum_{i\odd}
      \frac{\xi_i(t)}{\xi_k(t)}
      \left[\vphantom{\hat h(t)}
        \langle k, k\pr|\velel|i, i\pr\rangle
        - \langle k, k\pr|\velel|i\pr, i\rangle
      \right]
      - \ii\frac{\dot n_k(t)}{2 n_k(t)}. \label{diagonal-alpha-short-sum}
\end{align}
Recasting the sum in~\eqref{diagonal-alpha-short-sum} in the form
\begin{align}
  \frac{2}{n_k(t)}
  \sum_{i\odd}
  \ee^{\ii\left(\varphi_{i,0}-\varphi_{k,0}\right)}
  \sqrt{\frac{n_i(t)}{2}}
  \sqrt{\frac{n_k(t)}{2}}
  \left[\vphantom{\hat h(t)}
    \langle k, k\pr|\velel|i, i\pr\rangle
    - \langle k, k\pr|\velel|i\pr, i\rangle
  \right]
    &=
      \frac{2}{n_k(t)}
      \sum_{ijl}
      \tilde\gamma_{2,ijkl}(t)
      \langle\tilde k\tilde l|\velel|\tilde i\tilde j\rangle
\end{align}
and making use of  the analytically known expression for $\dot n_k(t)$ \cite{tdrnot},
\begin{align}
  \dot n_k(t)
    &=
      4\Im\sum_{ijl}
      \tilde\gamma_{2,ijkl}(t)
      \langle\tilde k\tilde l|\velel|\tilde i\tilde j\rangle,
\end{align}
gives
\begin{align}
  \alpha_{kk}(t)
  + \alpha_{k\pr k\pr}(t)
    &=
      \langle k|\hat h(t)|k\rangle
      + \langle k\pr|\hat h(t)|k\pr\rangle
      + \frac{2}{n_k(t)}
      \Re\sum_{ijl}
      \tilde\gamma_{2,ijkl}(t)
      \langle\tilde k\tilde l|\velel|\tilde i\tilde j\rangle. \label{sum-diagonal-alphas}
\end{align}
Equation~\eqref{sum-diagonal-alphas} reflects the freedom to distribute the global phase of $|i,i\pr\rangle-|i\pr,i\rangle$ in~\eqref{universal-expansion-xi} among orbital $i$ and orbital $i\pr$. Choosing
\begin{align}
  \alpha_{kk}(t)
    &=
      \alpha_{k\pr k\pr}(t)
      -\langle k\pr|\hat h(t)|k\pr\rangle
      +\langle k|\hat h(t)|k\rangle \label{eq:fix-new-phase-convention}
\end{align}
it is found that for both odd and even $k$ the final result reads
\begin{equation}
  \alpha_{kk}(t)
    =
      \frac{1}{n_k(t)}
      \left[
        \langle\tilde k|\hat h(t)|\tilde k\rangle
        + \Re\sum_{ijl}
        \tilde\gamma_{2,ijkl}(t)
        \langle\tilde k\tilde l|\velel|\tilde i\tilde j\rangle
      \right].
\end{equation}

\section{Variational determination of the spin-triplet ground state}\label{appendix:triplet-variation}
As in~\cite{tdrnot}, we define an energy functional $\tilde E$ taking into account the constraints
$
  \sum_i n_i
    = N=
      2$,
  $
  \langle i|j\rangle
    =
      \kronecker_{ij}$,
 $ n_i
    \geq
      0$,
  $
  n_i
    \leq
      1$ 
via the Lagrange parameters $\epsilon$ and $\lambda_{ij}$ as well as the Karush-Kuhn-Tucker parameters~\cite{giesbthesis, numerical-optimization} $\epsilon_i^0$ and $\epsilon_i^1$, respectively. Additionally, the degeneracy $n_i=n_{i\pr}$ is enforced via the Lagrange parameter $\epsilon_i^\mathrm{d}$ for odd $i$. The functional $\tilde E$ reads
\begin{align}
  \tilde E
    &=
      \sum_{i}
      \langle \tilde i|\hat h_0|\tilde i\rangle
      + \sum_{ijkl}
      \frac{
        \gaMM_{2,ijkl}
      }{
        \sqrt{\ntilde{i}\ntilde{k}}
      }
      \langle \tilde k\tilde l|\velel|\tilde i\tilde j\rangle\nonumber\\
    &\quad
      - \epsilon\left[\vphantom{\sum_i}
        \sum_i\ntilde{i}
        - 2
      \right]
      - \sum_i\sum_{j\neq i}
      \lambda_{ij}\nntilde{i}{j}
      - \sum_i\left[\vphantom{\sum_i}
        \epsilon_i^0\ntilde{i}
        + \epsilon_i^1\left(
          1-\ntilde{i}
        \right)
      \right]
      - \sum_{i\odd}
      \epsilon_i^\mathrm{d}\left[\vphantom{\sum_i}
        \ntilde{i}
        - \ntilde{{i\pr}}
      \right], \label{energy-functional-with-constraints}
\end{align}
where the slackness conditions~\cite{giesbthesis} are $0=\epsilon_i^0 n_i=\epsilon_i^1 (1-n_i)$, and 
\begin{align*}
  \gaMM_{2,ijkl}=
      (-1)^{i-k}
      \frac{
        \ee^{\ii\left(
          \varphi_{i,0}
          - \varphi_{k,0}
        \right)}
      }
      {
        2
      }
      \kronecker_{ij\pr}
      \kronecker_{kl\pr},
\end{align*}
which is a constant regarding the variation of RNOs. Variation of the energy functional~\eqref{energy-functional-with-constraints} with respect to $\langle\tilde m|$ and $|\tilde m\rangle$ yields
\begin{align}
  \epsilon_m |\tilde m\rangle
    &=
      \left\{
        \hat h_0
        +2\sum_{jl}
          \frac{
            \gaMM_{2,mjml}
          }{
            \ntilde{m}
          }
          \langle\tilde l|\velel|\tilde j\rangle(x)
          -\frac{
            1
          }{
            \ntilde{m}
          }\Re\left[\sum_{ijl}
            \frac{
              \gaMM_{2,ijml}
            }{
              \sqrt{\ntilde{i}\ntilde{m}}
            }
            \langle \tilde m\tilde l|\velel|\tilde i\tilde j\rangle
          \right]
      \right\}|\tilde m\rangle\nonumber\\
    &\qquad\qquad
      +\sum_{i\neq m}\left\{
        2\sum_{jl}
          \frac{
            \gaMM_{2,ijml}
          }{
            \sqrt{\ntilde{i}\ntilde{m}}
          }
          \langle\tilde l|\velel|\tilde j\rangle(x)
        -\lambda_{mi}
      \right\}|\tilde i\rangle \label{energy-variation-intermediate}
\end{align}
and its Hermitian conjugate, respectively. 
\end{widetext}
The orbital energies $\epsilon_m$ are defined as
\begin{align*}
  \epsilon_m
    &=
      \epsilon
      + \epsilon_m^0
      - \epsilon_m^1
      + \epsilon^\mathrm{d}_m\kronecker_{m\odd}
      - \epsilon^\mathrm{d}_{m-1}\kronecker_{m\even}.
\end{align*}
The phases $\varphi_{i,0}$ in~\eqref{eq:frozen-phase-factors} are defined such that the ground state NOs of the model system may be chosen real. Assuming real ground state NOs,~\eqref{energy-variation-intermediate} and its Hermitian conjugate yield
\begin{align*}
  \epsilon_k
    &=
      \frac{1}{n_k}
      \left[
        \langle\tilde k|\hat h_0|\tilde k\rangle
        + \sum_{ijl}\tilde\gamma_{2,ijkl}
        \langle\tilde k\tilde l|\velel|\tilde i\tilde j\rangle
      \right].
\end{align*}
For correlated systems, i.e., in general non-integer ONs, we have $0=\epsilon_i^0=\epsilon_i^1$ so that $\epsilon_k=\epsilon + \epsilon^\mathrm{d}_m\kronecker_{m\odd} - \epsilon^\mathrm{d}_{m-1}\kronecker_{m\even}$. Hence, each sum of two associated orbital energies in the ground state fulfills
\begin{align*}
  \epsilon_k
  + \epsilon_{k\pr}
    &=
      2 \epsilon.
\end{align*}
Moreover, the set of ground state RNOs is a stationary point of the imaginary-time propagation, as already pointed out for the singlet in~\cite{tdrnot}.

\section{Newton scheme for finding ground state ONs} \label{appendix:newton-derivation}
In this appendix, a scheme for finding the correct ground state ONs is presented when the phase convention of section~\ref{sec:former-phase-convention} is chosen. Section~\ref{sec:discussion-full-imaginary-time-propagation} contains a brief discussion why this ``tuning'' of ONs is necessary. 
The variational calculus in appendix~\ref{appendix:triplet-variation} shows that the converged RNOs associated with the correct ground state ONs fulfill
\begin{align}
  \epsilon_k
  + \epsilon_{k\pr}
    &=
      E \label{newton-derivation-start-criterion}
    =
      E_0.
\end{align}
For $N_\mathrm{o}$ RNOs, due to the pairwise degeneracy of ONs and the constraint $\sum_k n_k=2$, there are $(N_\mathrm{o}/2-1)$ free parameters. With
\begin{align}
  n_1 = n_2
    &=
      1
      - \sum_{\text{odd }i\neq 1} n_i \label{fix-first-occupation-number}
\end{align}
and 
\begin{align*}
  \bm{n}
    &=
      \left(
        n_3, n_5, \dots, n_{N_\mathrm{o} - 1}
      \right)\trans,\\
  \bm{F}(\bm{n})
    &=
      \left(
        F_3(\bm{n}), F_5(\bm{n}), \dots, F_{N_\mathrm{o} - 1}(\bm{n})
      \right)\trans,\\
  F_m(\bm{n})
    &=
      \epsilon_m + \epsilon_{m+1}
      - \epsilon_1 - \epsilon_2
\end{align*}
the root of $\bm{F}$ fulfills~\eqref{newton-derivation-start-criterion} for all $k$. We thus search the root of $\bm{F}$ using the Newton-Raphson scheme. One iteration step from configuration $\bm{n}^{(i)}$ to configuration $\bm{n}^{(i+1)}$ is performed according to
\begin{align*}
  \bm{J}\left(
    \bm{n}^{(i+1)}
    - \bm{n}^{(i)}
  \right)
    &=
      -\bm{F}(\bm{n}^{(i)})
\end{align*}
where $\bm{J}=\left[J_{mn}\right]=\left[\derivative_{n_n} F_m\right]$ (for odd $m\neq 1$ and odd $n\neq 1$) is the Jacobian matrix. The derivatives $\derivative_{n_n} F_m$ are calculated using
\begin{align}
  \derivative_{n_n}
  |\tilde m\rangle
    &=
      |m\rangle
      \derivative_{n_n}
      \sqrt{n_m}. \label{eq:newton-scheme-approximation}
\end{align}
In practice, also the converged NOs $|m_0(\bm n)\rangle$ for a given ON configuration $\bm n$ change if the ON $n_n$ (and thus also $n_1$, $n_2$, and $n_{n\pr}$) is modified. However, the approximation~\eqref{eq:newton-scheme-approximation} yields smooth convergence.
\begin{widetext}
Because of~\eqref{fix-first-occupation-number} $\derivative_{n_n} n_1 = -1$ for odd $n\neq 1$. Hence for odd $n\neq 1$
\begin{align*}
  \derivative_{n_n}\gamma_{2,ijkl}
    &=
      \left[
        \kronecker_{ni}
        + \kronecker_{nj}
        + \kronecker_{nk}
        + \kronecker_{nl}
      \right]\frac{\gamma_{2,ijkl}}{2 n_n}
      -\left[
        \kronecker_{1i}
        + \kronecker_{1j}
        + \kronecker_{1k}
        + \kronecker_{1l}
      \right]\frac{\gamma_{2,ijkl}}{2 n_1}.
\end{align*}
Assuming real NOs for the ground state one finds for odd $m\neq 1$
\begin{align*}
  F_m
    &=
      \langle m|\hat h_0|m\rangle
      + \langle m\pr|\hat h_0|m\pr\rangle
      + 2\sum_{ij}
      \frac{\gamma_{2,ijmm\pr}}{n_m}
      \langle m m\pr|\velel|i j\rangle
      - \langle 1|\hat h_0|1\rangle
      - \langle 2|\hat h_0|2\rangle
      - 2\sum_{ij}
      \frac{\gamma_{2,ij12}}{n_1}
      \langle 12|\velel|ij\rangle.
\end{align*}
The phases $\varphi_i$ in $\gamma_{2,ijkl}$ can be set to the frozen phases $\varphi_{i,0}$ of the PINO phase convention~\eqref{eq:frozen-phase-factors} because the time-independent ground state is sought. For odd $n\not\in\{1,m\}$ follows
\begin{align*}
  \derivative_{n_n}
  F_m
    &=
      \frac{1}{2(n_n n_m)^{3/2}}
      \langle\tilde m\tilde m\pr|\velel\left[
        |\tilde n\tilde n\pr\rangle
        - |\tilde n\pr\tilde n\rangle
      \right]
      - \frac{\ee^{\ii\varphi_{m,0}}}{2(n_1 n_m)^{3/2}}
      \langle\tilde m\tilde m\pr|\velel\left[
        |\tilde1\tilde2\rangle
        -|\tilde2\tilde1\rangle
      \right]\nonumber\\
    &\qquad
      - \frac{\ee^{\ii\varphi_{n,0}}}{2(n_n n_1)^{3/2}}
      \langle\tilde1\tilde2|\velel\left[
        |\tilde n\tilde n\pr\rangle
        - |\tilde n\pr\tilde n\rangle
      \right]
      - \sum_{\text{odd } i\neq 1}
      \frac{\ee^{\ii\varphi_{i,0}}}{2 \sqrt{n_1^5 n_i}}
      \langle\tilde1\tilde2|\velel\left[
        |\tilde i\tilde i\pr\rangle-
        |\tilde i\pr\tilde i\rangle
      \right],
\end{align*}
for the diagonal element 
\begin{align*}
  \derivative_{n_m}
  F_m
    &=
      - \sum_{\text{odd } i\neq m}
      \frac{\ee^{\ii\left[\varphi_{m,0} + \varphi_{i,0}\right]}}{2 \sqrt{n_m^5 n_i}}
      \langle \tilde m\tilde m\pr|\velel\left[
        |\tilde i\tilde i\pr\rangle
        - |\tilde i\pr\tilde i\rangle
      \right]
      - \sum_{\text{odd } i\neq 1}
      \frac{\ee^{\ii\varphi_{i,0}}}{2 \sqrt{n_1^5 n_i}}
      \langle\tilde1\tilde2|\velel\left[
        |\tilde i\tilde i\pr\rangle
        - |\tilde i\pr\tilde i\rangle
      \right]\nonumber\\
    &\qquad
      - \frac{\ee^{\ii\varphi_{m,0}}}{(n_1 n_m)^{3/2}}
      \langle \tilde m\tilde m\pr|\velel\left[
        |\tilde1\tilde2\rangle
        - |\tilde2\tilde1\rangle
      \right].
\end{align*}

\label{appendix:end}

\end{widetext}


\begin{thebibliography}{13}%
\makeatletter
\providecommand \@ifxundefined [1]{%
 \@ifx{#1\undefined}
}%
\providecommand \@ifnum [1]{%
 \ifnum #1\expandafter \@firstoftwo
 \else \expandafter \@secondoftwo
 \fi
}%
\providecommand \@ifx [1]{%
 \ifx #1\expandafter \@firstoftwo
 \else \expandafter \@secondoftwo
 \fi
}%
\providecommand \natexlab [1]{#1}%
\providecommand \enquote  [1]{``#1''}%
\providecommand \bibnamefont  [1]{#1}%
\providecommand \bibfnamefont [1]{#1}%
\providecommand \citenamefont [1]{#1}%
\providecommand \href@noop [0]{\@secondoftwo}%
\providecommand \href [0]{\begingroup \@sanitize@url \@href}%
\providecommand \@href[1]{\@@startlink{#1}\@@href}%
\providecommand \@@href[1]{\endgroup#1\@@endlink}%
\providecommand \@sanitize@url [0]{\catcode `\\12\catcode `\$12\catcode
  `\&12\catcode `\#12\catcode `\^12\catcode `\_12\catcode `\%12\relax}%
\providecommand \@@startlink[1]{}%
\providecommand \@@endlink[0]{}%
\providecommand \url  [0]{\begingroup\@sanitize@url \@url }%
\providecommand \@url [1]{\endgroup\@href {#1}{\urlprefix }}%
\providecommand \urlprefix  [0]{URL }%
\providecommand \Eprint [0]{\href }%
\providecommand \doibase [0]{http://dx.doi.org/}%
\providecommand \selectlanguage [0]{\@gobble}%
\providecommand \bibinfo  [0]{\@secondoftwo}%
\providecommand \bibfield  [0]{\@secondoftwo}%
\providecommand \translation [1]{[#1]}%
\providecommand \BibitemOpen [0]{}%
\providecommand \bibitemStop [0]{}%
\providecommand \bibitemNoStop [0]{.\EOS\space}%
\providecommand \EOS [0]{\spacefactor3000\relax}%
\providecommand \BibitemShut  [1]{\csname bibitem#1\endcsname}%
\let\auto@bib@innerbib\@empty

\bibitem{scrinzi} A.\ Scrinzi, in {\em Attosecond and XUV Physics} edited by Th.\ Schultz, M.\ Vrakking (Wiley-VCH, Weinheim, 2014), p.~257--292.
\bibitem{coulson} C.A.\ Coulson, \rmp {\bf 32}, 170 (1960).
\bibitem [{\citenamefont {Runge}\ and\ \citenamefont
  {Gross}(1984)}]{runge-gross}%
  \BibitemOpen
  \bibfield  {author} {\bibinfo {author} {\bibfnamefont {E.}~\bibnamefont
  {Runge}}\ and\ \bibinfo {author} {\bibfnamefont {E.~K.~U.}\ \bibnamefont
  {Gross}},\ }\href {\doibase 10.1103/PhysRevLett.52.997} {\bibfield  {journal}
  {\bibinfo  {journal} {Phys. Rev. Lett.}\ }\textbf {\bibinfo {volume} {52}},\
  \bibinfo {pages} {997} (\bibinfo {year} {1984})}\BibitemShut {NoStop}%
\bibitem{ullrich} C.A.\ Ullrich, {\em Time-Dependent Density Functional Theory, Concepts and Applications}, (Oxford University Press, Oxford, 2012).
\bibitem [{\citenamefont {Helbig}\ \emph {et~al.}(2011)\citenamefont {Helbig},
  \citenamefont {Fuks}, \citenamefont {Tokatly}, \citenamefont {Appel},
  \citenamefont {Gross},\ and\ \citenamefont {Rubio}}]{helbig}%
  \BibitemOpen
  \bibfield  {author} {\bibinfo {author} {\bibfnamefont {N.}~\bibnamefont
  {Helbig}}, \bibinfo {author} {\bibfnamefont {J.}~\bibnamefont {Fuks}},
  \bibinfo {author} {\bibfnamefont {I.}~\bibnamefont {Tokatly}}, \bibinfo
  {author} {\bibfnamefont {H.}~\bibnamefont {Appel}}, \bibinfo {author}
  {\bibfnamefont {E.}~\bibnamefont {Gross}}, \ and\ \bibinfo {author}
  {\bibfnamefont {A.}~\bibnamefont {Rubio}},\ }\href {\doibase
  http://dx.doi.org/10.1016/j.chemphys.2011.06.010} {\bibfield  {journal}
  {\bibinfo  {journal} {Chemical Physics}\ }\textbf {\bibinfo {volume} {391}},\
  \bibinfo {pages} {1 } (\bibinfo {year} {2011})}\BibitemShut {NoStop}%
\bibitem{petersilka} M.\ Petersilka, E.K.U.\ Gross, Laser Phys.\ {\bf 9}, 105 (1999).
\bibitem{wilken} F.\ Wilken, D.\ Bauer, \prl {\bf 97}, 203001 (2006).
\bibitem{wilken2} F.\ Wilken, D.\ Bauer, \pra {\bf 76}, 023409 (2007). 



\bibitem{colemanyukalov} A.J.\ Coleman, V.I.\ Yukalov, {\em Reduced Density Matrices, Coulson's Challenge}, Lecture Notes in Chemistry 72,  (Springer, Berlin Heidelberg, 2000).
\bibitem{cios} J.\ Cioslowski (Ed.), {\em Many-electron densities and reduced density matrices}, Mathematical and Computational Chemistry Series, (Kluwer/Plenum, New York, 2000).
\bibitem{gido} N.I.\ Gidopoulos, S.\ Wilson (Eds.), {\em Electron Density, Density Matrix and Density Functional Theory in Atoms, Molecules and the Solid State}, Progress in Theoretical Chemistry and Physics, (Kluwer, Dordrecht, 2003).
\bibitem{mazz} D.A.\ Mazziotti (Ed.), {\em Reduced-Density-Matrix Mechanics}, Advances in Chemical Physics Vol.\ 134, (Wiley, Hoboken, 2007).
\bibitem{mazz2} D.A.\ Mazziotti, Chem.\ Rev.\ {\bf 112}, 244 (2012).
\bibitem{pernalTD} K.\ Pernal, O.\ Gritsenko, E.J.\ Baerends, \pra {\bf 75}, 012506 (2007)
\bibitem{appelthesis} H.\ Appel, {\em Time-Dependent Quantum Many-Body Systems: Linear Response, Electronic Transport, and Reduced Density Matrices}, (Doctoral Thesis, Free University Berlin, 2007); http://www.diss.fu-berlin.de/diss/receive/FUDISS\_thesis\_000000003068
\bibitem{gies1} K.J.H.\ Giesbertz, E.J.\ Baerends, O.V.\ Gritsenko, \prl {\bf 101}, 033004 (2008).
\bibitem{giesbthesis} K.J.H.\ Giesbertz, {\em Time-Dependent One-Body Reduced Density Matrix Functional Theory, Adiabatic Approximations and Beyond}, (PhD Thesis, Free University Amsterdam, 2010); http://dare.ubvu.vu.nl/handle/1871/16289

\bibitem{requist} R.\ Requist, O.\ Pankratov, \pra {\bf 81}, 042519 (2010).
\bibitem{appelgross} H.\ Appel, E.K.U.\ Gross, Europhys.\ Lett.\ {\bf 92}, 23001 (2010).

\bibitem{loewdin1} P.-O.\ L\"owdin, Phys.\ Rev.\ {\bf 97}, 1474 (1955).
\bibitem{loewdintwoelecs} P.-O.\ L\"owdin, H.\ Shull, Phys.\ Rev.\ {\bf 101}, 1730 (1956).

\bibitem{giesbnobasis2014} K.J.H.~Giesbertz, Chem.~Phys.~Lett.~{\bf 591}, 220 (2014)

\bibitem{model1}  R.\ Grobe and J.H.\ Eberly,  \prl {\bf 68}, 2905 (1992); S.L.\ Haan, R.\ Grobe, and J.H.\ Eberly, \pra {\bf 50}, 378 (1994). 
\bibitem{model2}  D.\ Bauer, \pra {\bf 56}, 3028 (1997); D.G.\ Lappas and R.\ van Leeuwen, J.\ Phys.\ B: At.\ Mol.\ Opt.\ Phys.\ {\bf 31}, L249 (1998); D.\ Bauer, F.\ Ceccherini,  \pra {\bf 60}, 2301 (1999); M.\ Lein, E.K.U.\ Gross, and V.\ Engel, \prl {\bf 85}, 4707 (2000); M.\ Thiele, E.K.U.\ Gross, and S.\ K\"ummel, \prl {\bf 100}, 153004 (2008).
\bibitem [{\citenamefont {Brics}\ and\ \citenamefont {Bauer}(2013)}]{tdrnot}%
  \BibitemOpen
  \bibfield  {author} {\bibinfo {author} {\bibfnamefont {M.}~\bibnamefont
  {Brics}}\ and\ \bibinfo {author} {\bibfnamefont {D.}~\bibnamefont {Bauer}},\
  }\href {\doibase 10.1103/PhysRevA.88.052514} {\bibfield  {journal} {\bibinfo
  {journal} {Phys. Rev. A}\ }\textbf {\bibinfo {volume} {88}},\ \bibinfo
  {pages} {052514} (\bibinfo {year} {2013})}\BibitemShut {NoStop}%
\bibitem{krueger} A.J.\ Krueger and N.T.\ Maitra, Phys.\ Chem.\ Chem.\ Phys.\ {\bf 11}, 4655 (2009).

\bibitem{giesbgritsbaer} K.J.H.\ Giesbertz, O.V.\ Gritsenko, and E.J.\ Baerends, J.\ Chem.\ Phys.\ {\bf 136}, 094104 (2012).
\bibitem{Meer} R.\ van Meer,  O.V.\ Gritsenko, K.J.H.\ Giesbertz, and E.J.\ Baerends,
 J.\ Chem.\ Phys.\ {\bf 138}, 094114 (2013).
\bibitem [{\citenamefont {Hochstuhl}\ \emph {et~al.}(2010)\citenamefont
  {Hochstuhl}, \citenamefont {Bauch},\ and\ \citenamefont {Bonitz}}]{mctdhf}%
  \BibitemOpen
  \bibfield  {author} {\bibinfo {author} {\bibfnamefont {D.}~\bibnamefont
  {Hochstuhl}}, \bibinfo {author} {\bibfnamefont {S.}~\bibnamefont {Bauch}}, \
  and\ \bibinfo {author} {\bibfnamefont {M.}~\bibnamefont {Bonitz}},\ }\href
  {http://stacks.iop.org/1742-6596/220/i=1/a=012019} {\bibfield  {journal}
  {\bibinfo  {journal} {Journal of Physics: Conference Series}\ }\textbf
  {\bibinfo {volume} {220}},\ \bibinfo {pages} {012019} (\bibinfo {year}
  {2010})}\BibitemShut {NoStop}%
\bibitem [{\citenamefont {Ruggenthaler}\ and\ \citenamefont
  {Bauer}(2009)}]{rabi-1}%
  \BibitemOpen
  \bibfield  {author} {\bibinfo {author} {\bibfnamefont {M.}~\bibnamefont
  {Ruggenthaler}}\ and\ \bibinfo {author} {\bibfnamefont {D.}~\bibnamefont
  {Bauer}},\ }\href {\doibase 10.1103/PhysRevLett.102.233001} {\bibfield
  {journal} {\bibinfo  {journal} {Phys. Rev. Lett.}\ }\textbf {\bibinfo
  {volume} {102}},\ \bibinfo {pages} {233001} (\bibinfo {year}
  {2009})}\BibitemShut {NoStop}%
\bibitem [{\citenamefont {Horn}\ and\ \citenamefont
  {Johnson}(2012)}]{matrixAnalysis}%
  \BibitemOpen
  \bibfield  {author} {\bibinfo {author} {\bibfnamefont {R.}~\bibnamefont
  {Horn}}\ and\ \bibinfo {author} {\bibfnamefont {C.}~\bibnamefont {Johnson}},\
  }\href@noop {} {\emph {\bibinfo {title} {Matrix Analysis}}},\ \bibinfo
  {edition} {2nd}\ ed.\ (\bibinfo  {publisher} {Cambridge University Press},\
  \bibinfo {address} {New York},\ \bibinfo {year} {2012})\BibitemShut {NoStop}%
\bibitem [{\citenamefont {Nocedal}\ and\ \citenamefont
  {Wright}(2006)}]{numerical-optimization}%
  \BibitemOpen
  \bibfield  {author} {\bibinfo {author} {\bibfnamefont {J.}~\bibnamefont
  {Nocedal}}\ and\ \bibinfo {author} {\bibfnamefont {S.}~\bibnamefont
  {Wright}},\ }\href@noop {} {\emph {\bibinfo {title} {Numerical
  Optimization}}},\ \bibinfo {edition} {2nd}\ ed.,\ Springer Series in
  Operations Research and Financial Engineering\ (\bibinfo  {publisher}
  {Springer},\ \bibinfo {address} {New York},\ \bibinfo {year}
  {2006})\BibitemShut {NoStop}%
\end{thebibliography}
\end{document}